\documentstyle[aps,epsfig]{revtex}

\date{January 2000}

        \newcommand{\be}{\begin{equation}}
        \newcommand{\ee}{\end{equation}}
        \newcommand{\bea}{\begin{eqnarray}}
        \newcommand{\eea}{\end{eqnarray}}
        \newcommand{\ban}{\begin{eqnarray*}}
        \newcommand{\ean}{\end{eqnarray*}}
        \newcommand{\half}{\frac{1}{2}}

\begin{document}

\begin{center}
{\Large\bf Chiral Symmetry Restoration at Nonzero Temperature in 
the $SU(3)_{r} \times SU(3)_{\ell}$ Linear Sigma Model}
\\[1cm]
Jonathan T.\ Lenaghan\footnote{Current address: Physics Department,
Brookhaven National Laboratory, Upton, NY 11973, USA}
\\ ~~ \\
{\it Physics Department, Yale University} \\
{\it New Haven, CT 06520, USA}
\\ ~~ \\ 
Dirk H.\ Rischke, J\"urgen Schaffner-Bielich
\\ ~~ \\ 
{\it RIKEN-BNL Research Center, Physics Department} \\
{\it Brookhaven National Laboratory, Upton, NY 11973, USA}
\\ ~~ \\ ~~ \\
\end{center}

\begin{abstract}
We study patterns of chiral symmetry breaking at zero temperature
and its subsequent restoration at nonzero temperature within 
the $SU(3)_{r} \times SU(3)_{\ell}$ linear sigma model. 
Gap equations for the masses of the scalar and pseudoscalar 
mesons and the non-strange and strange quark condensates are
systematically derived in the Hartree approximation via
the Cornwall--Jackiw--Tomboulis formalism. 
In the chiral limit, the chiral symmetry restoring transition
is found to be first order, as predicted by universality arguments.
Taking the experimental values for the 
meson masses, however, the transition
is crossover. The absence of the $U(1)_A$ anomaly is found to
drive this transition closer to being first order.
At large temperatures, the mixing angles between octet and singlet 
states approach ideal flavor mixing. 
\end{abstract}

\section{Introduction}

Chiral symmetry is broken in the vacuum of quantum chromodynamics (QCD).  
At temperatures of about 150 MeV, lattice QCD calculations 
indicate that chiral symmetry is restored \cite{lattice}.  
The order of the phase transition 
depends on the mass of the non-strange up and down quarks, 
$m_u \simeq m_d$, and the mass of the strange 
quark, $m_s$ \cite{Columbia}. In nature, $m_u \simeq m_d \sim 10$
MeV and $m_s \sim 100$ MeV \cite{msubs}.  At temperatures 
on the order of 150 MeV, heavier quark flavors do not 
play an essential role.

For $N_f$ massless quark flavors, the QCD Lagrangian has a
$SU(N_f)_r \times SU(N_f)_{\ell} \times U(1)_{A}$ symmetry. 
In the vacuum, a non-vanishing expectation value of the 
quark condensate, $\langle \bar{q}_{\ell} \, q_{r} \rangle \neq 0$, 
spontaneously breaks this symmetry to the diagonal
$SU(N_f)_{V}$ group of vector transformations, $V=r+\ell$. 
For $N_f = 3$, the effective, low-energy degrees of 
freedom of QCD are the scalar and pseudoscalar mesons.
Since mesons are $q \bar{q}$ states, they 
fall in singlet and octet representations of $SU(3)_{V}$.

The $SU(N_f)_r \times SU(N_f)_{\ell} \times U(1)_A$ symmetry 
of the QCD Lagrangian is also explicitly broken by nonzero quark 
masses. 
For $M \leq N_{f}$ degenerate 
quark flavors, a $SU(M)_{V}$ symmetry is preserved.  
If $M < N_{f}$, the mass eigenstates are mixtures of 
singlet and octet states.
For instance, in the pseudoscalar meson sector, this 
mixing occurs between the $\eta$ and the $\eta'$ meson, with 
the $\eta$ meson being mostly octet and the $\eta'$ meson being mostly 
singlet, with a mixing angle of about -10 to -20 degrees
\cite{PDB}. 

As shown by 't Hooft \cite{tHooft}, instantons also break 
the $U(1)_{A}$ symmetry explicitly to $Z(N_{f})_{A}$ 
\cite{PisarskiWilczek}.  
For the low-energy dynamics of QCD, however, 
this discrete symmetry is irrelevant.

Pisarski and Wilczek \cite{PisarskiWilczek} discussed the
order of the chiral transition using renormalization group
arguments in the framework of the linear sigma model. 
This model is the effective theory for the low-energy degrees of
freedom of QCD and incorporates the global 
$SU(N_f)_r \times SU(N_f)_{\ell} \times U(1)_A$ symmetry, but
not the local $SU(3)_c$ color symmetry. They found that,
for $N_f=2$ flavors of massless quarks, the 
transition can be of second order, if the $U(1)_A$ symmetry
is explicitly broken by instantons. It is driven first order by fluctuations, 
if the $U(1)_A$ symmetry is restored at $T_c$.
For $N_f=3$ massless flavors, the transition is always first order.
In this case, the term which breaks the $U(1)_A$ symmetry
explicitly is a cubic
invariant, and consequently drives the transition first order. 
In the absence of explicit $U(1)_A$ symmetry breaking, the transition is 
fluctuation-induced of first order.

In nature, the chiral symmetry of QCD is explicitly
broken by nonzero quark masses.
In this case, one has to resort to numerical
calculations to determine the order of the chiral transition. 
At present, however, lattice QCD data have not unambiguously 
settled this issue. For physical values of the quark masses, 
calculations with staggered fermions \cite{Columbia} 
favor a smooth crossover transition, while calculations
with Wilson fermions \cite{Wilson} predict the transition to be first order.

As an alternative to lattice QCD calculations, one
can also use the linear sigma model
to make predictions on the order of the phase transition in QCD.
Furthermore, various symmetry-breaking scenarios can be
more easily investigated than on the lattice.
Studying the linear sigma model at nonzero temperature, however,
requires many-body resummation schemes,
because infrared divergences cause naive perturbation theory 
to break down \cite{Dolan}.

For $N_{f} = 2$, effective chiral models for QCD have 
been studied extensively, because in this case 
$SU(2)_{r} \times SU(2)_{\ell}$ is isomorphic to 
$O(4)$, and $O(N)$ models \cite{GellMann} in general 
are particularly amenable to many-body 
approximations at nonzero temperature. For an incomplete 
list of references, see \cite{JDRenorm}, where some of us 
(J.T.L.\ and D.H.R.) studied the $O(N)$ model in the
Hartree and large--$N$ approximations.

For $N_{f} =3$, many-body resummation schemes 
become considerably more involved due to the larger 
number of degrees of freedom. 
The $SU(3)_{r} \times SU(3)_{\ell}$ linear sigma model \cite{Levy}
was previously studied at nonzero temperature in \cite{su3T}.
The genuine problem of these approaches is that they employ
methods related to the standard loop expansion to
compute the effective potential. At nonzero temperature,
the loop expansion is known to fail
in the case of spontaneously broken symmetry, since it
generates imaginary masses for the particles.
The physical reason for this failure is that contributions from
thermal excitations to the masses are neglected.
This can be amended by self-consistent resummation schemes like
the mean-field or the Hartree approximation.
In \cite{Jurgen1}, one of us (J.S.-B.) studied the linear sigma
model in the mean-field approximation. 

In this work, we study the
$SU(3)_{r} \times SU(3)_{\ell}$ linear sigma model in the
Hartree approximation.  We derive this approximation 
systematically from the  
Cornwall--Jackiw--Tomboulis (CJT) formalism \cite{CJT}.  
Thus, we extend previous work for $N_f=2$ \cite{JDRenorm} to 
$N_f=3$.  We study  possible patterns of symmetry 
breaking in the vacuum
and its subsequent restoration at nonzero temperature. 
We focus on both the cases where chiral symmetry is 
and where it is not explicitly broken by nonzero quark masses.
Lattice QCD data indicate that the $U(1)_A$ anomaly becomes 
small near the chiral transition \cite{latticeu1a}.   Therefore,
we also study the influence of the $U(1)_A$ anomaly 
on symmetry restoration at nonzero temperature.

This paper is organized as follows.  In Sec.\ \ref{II}, we introduce 
the $SU(3)_{r} \times SU(3)_{\ell}$ linear sigma model.  In Sec.\ 
\ref{III}, the possible patterns of symmetry breaking
in the vacuum are discussed. 
Section \ref{IV} is devoted to the vacuum properties 
of the model at tree level.  In Sec.\ \ref{V}, we derive the 
gap equations for the condensates and the masses in Hartree
approximation via the CJT effective potential.  Numerical results 
are presented in Sec.\ \ref{VI}. 
We conclude this work in Sec.\ \ref{VII} with a summary of 
our results. 

We use the imaginary-time formalism to compute quantities at
nonzero temperature. Our notation is 
\be 
\int_k \, f(k) \equiv T \sum_{n=-\infty}^{\infty} 
                       \int \frac{d^{3}k}{(2\pi)^{3}} \,
         f(2 \pi i n T,{\bf k}) \,\,\,\, , \,\,\,\,
\int_x \, f(x) \equiv \int^{1/T}_{0} d \tau \int d^{3}{\bf x} \,
f(\tau,{\bf x}) \,\, .
\ee 
We use units $\hbar=c=k_{B}=1$.  The metric tensor is $g^{\mu \nu}
= {\rm diag}(+,-,-,-)$.  Throughout this work,  
all latin subscripts are adjoint $U(3)$ indices, $a=0,\ldots,8$, and 
a summation over repeated indices is understood.

\section{The Linear Sigma Model for Three Flavors} \label{II}

The Lagrangian of the $SU(3)_{r} \times SU(3)_{\ell}$ 
linear sigma model is given by \cite{GellMann}
\bea \label{L}
{\cal L}(\Phi) &=& 
{\rm Tr}  \left( \partial_{\mu} \Phi^{\dagger} 
\partial^{\mu} \Phi 
-  m^2 \, \Phi^{\dagger} 
\Phi \right) - 
\lambda_{1} \left[ {\rm Tr}  \left( \Phi^{\dagger}
 \Phi  \right) \right]^{2} - 
\lambda_{2} {\rm Tr}  \left( \Phi^{\dagger} 
 \Phi  \right)^{2} \nonumber \\ 
&+& c \left[ {\rm Det} \left( \Phi \right) + 
{\rm Det}  \left( \Phi^{\dagger} \right) \right]  
+ {\rm Tr} \left[H  (\Phi + \Phi^{\dagger})\right] \,\, . 
\eea
$\Phi$ is a complex $3 \times 3$ matrix parametrizing the
scalar and pseudoscalar meson nonets,
\begin{mathletters}
\be
\Phi =T_{a} \, \phi_{a} =   T_{a} \, (\sigma_{a} + 
        i \pi_{a})\,\, , \label{defphi}
\ee
where $\sigma_{a}$ are the scalar fields and
$\pi_{a}$ are the pseudoscalar fields.
The $3 \times 3$ matrix $H$ breaks the symmetry explicitly and is chosen as
\be
H = T_{a} \, h_{a} \,\, , 
\ee
\end{mathletters}
where $h_{a}$ are nine external fields. 
$T_{a} = \hat{\lambda}_{a}/2$ are the generators of 
$U(3)$, where $\hat{\lambda}_{a}$ are the Gell-Mann matrices with 
$\hat{\lambda}_{0} = \sqrt{\frac{2}{3}} \, {\bf 1}$.  The $T_{a}$ are 
normalized such that 
${\rm Tr} (T_{a} T_{b}) = \delta_{ab}/2$ and
obey the $U(3)$ algebra with
\begin{mathletters}
\bea
\left[T_{a},T_{b}\right] &=& i \, f_{abc} \, T_{c} \,\, , \\
\left\{T_{a},T_{b}\right\} &=&  d_{abc} \, T_{c} \,\, ,
\eea
where $f_{abc}$ and $d_{abc}$ for $a,b,c=1,\ldots,8$
are the standard antisymmetric and symmetric structure 
constants of $SU(3)$ and 
\be     
f_{ab0} \equiv 0 \,\,\,\,  ,\,\,\,\,
d_{ab0} \equiv \sqrt{\frac{2}{3}} \, \delta_{ab} \,\, .
\ee
\end{mathletters}

In Eq.\ (\ref{L}), $m^{2}$ is the tree-level 
mass of the fields in the absence of symmetry 
breaking, $\lambda_{1}$ and $\lambda_{2}$ are 
the two possible quartic coupling constants, and 
$c$ is the cubic coupling constant.  In four dimensions, 
the cubic and the two quartic terms are the only 
relevant $SU(3)_{r} \times SU(3)_{\ell}$ invariant
operators.

The terms in the first line of Eq.\ (\ref{L}) are actually invariant under 
the larger group of 
$U(3)_{r} \times U(3)_{\ell}$ symmetry transformations, 
\be \label{trans}
\Phi \longrightarrow U_{r} \, \Phi \, U_{\ell}^{\dagger} \,\,\, ,
\,\,\,\, U_{r,\ell} \equiv 
\exp\left(i \, \omega_{r,\ell}^{a} \, T^{a}\right) \,\, .
\ee
Introducing $\omega_{V,A}^{a} \equiv 
\left( \omega_{r}^{a} \pm \omega_{\ell}^{a} \right) /2$,
the right- and left-handed symmetry transformations 
can be alternatively written as vector, $V = r + \ell$, and 
axial vector, $A = r - \ell$, transformations.   It is then 
obvious that $\Phi$ is a singlet under $U(1)_{V}$ transformations 
$\exp(i \, \omega_{V}^{0} T^{0})$. This $U(1)_{V}$ is
the $U(1)$ of baryon number conservation and thus always respected.
The terms in the first line of Eq.\ (\ref{L}) are therefore invariant 
under $SU(3)_{r} \times SU(3)_{\ell} \times U(1)_{A} \cong
SU(3)_{V} \times U(3)_{A}$.

The determinant terms correspond to the $U(1)_A$ anomaly
in the QCD vacuum. As shown by 't Hooft \cite{tHooft},
they arise from instantons. These terms are invariant under 
$SU(3)_{r} \times SU(3)_{\ell} \cong SU(3)_{V} \times SU(3)_{A}$ 
transformations, but break the $U(1)_{A}$ symmetry of the 
Lagrangrian explicitly. 
The last term in Eq.\ (\ref{L}) breaks the axial and possibly the $SU(3)_{V}$
vector symmetries explicitly. The patterns of 
explicit symmetry breaking will be discussed in detail in Sec.\ \ref{III}.

The $\sigma_{a}$ fields are members of the 
scalar ($J^{P} = 0^{+}$) nonet and the $\pi_{a}$ 
fields are members of the pseudoscalar ($J^{P} = 0^{-}$) nonet, 
\begin{mathletters}
\bea 
T_{a} \, \sigma_{a} &=& \frac{1}{\sqrt{2}} \left( \begin{array}{ccc} 
        \frac{1}{\sqrt{2}} \, a_{0}^{0} + 
        \frac{1}{\sqrt{6}} \, \sigma_{8} +
        \frac{1}{\sqrt{3}} \, \sigma_{0} & a_{0}^{-} & \kappa^{-} \\
        a_{0}^{+} & -\frac{1}{\sqrt{2}} \, a_{0}^{0} +
        \frac{1}{\sqrt{6}} \, \sigma_{8} +
        \frac{1}{\sqrt{3}} \, \sigma_{0} & \bar{\kappa}^{0} \\
        \kappa^{+} & \kappa^{0} & -\frac{2}{\sqrt{3}} \, \sigma_{8} +
        \frac{1}{\sqrt{3}} \, \sigma_{0} \end{array} \right) \,\, , \\
T_{a} \, \pi_{a} &=& \frac{1}{\sqrt{2}} \left( \begin{array}{ccc} 
        \frac{1}{\sqrt{2}} \, \pi^{0} + \frac{1}{\sqrt{6}} \, \pi_{8} +
        \frac{1}{\sqrt{3}} \, \pi_{0} & \pi^{-} & K^{-} \\
        \pi^{+} & -\frac{1}{\sqrt{2}} \, \pi^{0} +
        \frac{1}{\sqrt{6}} \, \pi_{8} +
        \frac{1}{\sqrt{3}} \, \pi_{0} & \bar{K}^{0} \\
        K^{+} & K^{0} & -\frac{2}{\sqrt{3}} \, \pi_{8} +
        \frac{1}{\sqrt{3}} \, \pi_{0} \end{array} \right) \,\, .
\eea
\end{mathletters}  
Here, $\pi^{\pm} \equiv (\pi_{1} \pm i \, \pi_{2})/\sqrt{2}$ 
and $\pi^{0} \equiv \pi_{3}$ are the charged and neutral 
pions, respectively.  $K^{\pm} \equiv (\pi_{4} \pm i \, 
\pi_{5})/\sqrt{2}$, $K^{0} \equiv (\pi_{6} + i \, \pi_{7})/\sqrt{2}$,
and $\bar{K}^{0} \equiv (\pi_{6} - i \, \pi_{7})/\sqrt{2}$ are 
the kaons.  In general, because the strange quark is much 
heavier than the up or down quarks, the $\pi_{0}$ and 
the $\pi_{8}$ are admixtures of the $\eta$ and the $\eta'$ meson.

The situation with the scalar nonet is not as clear.
The parity partner of the pion is the $a_{0}(980)$ meson, i.e.,
$a_{0}^{\pm} \equiv (\sigma_{1} \pm i \, \sigma_{2})/\sqrt{2}$ 
and $a_{0}^{0} \equiv \sigma_{3}$. 
We identify the parity partner of the kaon with the $\kappa$ 
meson [now referred to as $K_{0}^{*}(1430)$ in \cite{PDB}].
Finally, in general the $\sigma_{0}$ and the $\sigma_{8}$ are admixtures 
of the $\sigma$ [now also referred to as 
$f_{0}(400-1200)$] and $f_{0}(1370)$ mesons.
[Instead of the $f_0(1370)$ meson, one could have chosen
the $f_0(980)$ meson. However, as we shall see in Sec.\ \ref{IV},
in the linear sigma model the mass of this state is closer to the $f_0(1370)$.]
For details concerning the phenomenological status of the scalar 
nonet, see, for instance \cite{scalarclass}.

In principle, there is the possibility that the scalar particles
are not diquark states, but formed from two quarks and two
antiquarks \cite{Jaffe}. Then we would associate the $\kappa$ with
the $\kappa(900)$ discovered in $\pi K$ scattering \cite{k900}.
Determining the vacuum properties of the linear sigma model, cf.\ Sec.\ 
\ref{IV}, the mass of the $\kappa$ turns out to be between
that of the $\kappa(900)$ and the $K_0^*(1430)$, 
while its width \cite{jj} is closer
to the observed width of the $\kappa(900)$.

Symmetry breaking gives the $\Phi$ field a vacuum 
expectation value, 
\be
\langle \Phi \rangle \equiv T_a \, \bar{\sigma}_{a} \,\, .
\ee
Shifting the $\Phi$ field by this vacuum expectation value,
the Lagrangian can be rewritten as \cite{ChanHay}, 
\bea
{\cal L}  &=& \frac{1}{2} \left[ \partial_{\mu} \sigma_{a} 
        \partial^{\mu} \sigma_{a} + \partial_{\mu} \pi_{a}
        \partial^{\mu} \pi_{a} - \sigma_{a} (m_{S}^{2})_{ab} 
        \sigma_{b} - \pi_{a} (m_{P}^{2})_{ab} 
        \pi_{b} \right]  \nonumber \\
         &+& \left({\cal G}_{abc}-\frac{4}{3} \, {\cal F}_{abcd} \,
        \bar{\sigma}_{d}\right)  \, \sigma_{a} \sigma_{b} \sigma_{c} - 
        3  \, \left({\cal G}_{abc} + \frac{4}{3}  \, {\cal H}_{abcd} 
         \, \bar{\sigma}_{d} \right)
          \, \pi_{a} \pi_{b} \sigma_{c} \nonumber \\
        &-& 2  \, {\cal H}_{abcd}  \, \sigma_{a} \sigma_{b}
        \pi_{c} \pi_{d} - \frac{1}{3}  \, {\cal F}_{abcd}  \, (
        \sigma_{a} \sigma_{b} \sigma_{c} \sigma_{d} + 
        \pi_{a} \pi_{b} \pi_{c} \pi_{d} ) 
        - U(\bar{\sigma}) \,\, ,
\eea
where 
\be \label{Utree}
U(\bar{\sigma}) = \frac{m^2}{2} \, \bar{\sigma}_{a}^{2} - 
        {\cal G}_{abc} \bar{\sigma}_{a} \bar{\sigma}_{b} 
        \bar{\sigma}_{c} + \frac{1}{3} {\cal F}_{abcd} 
        \bar{\sigma}_{a} \bar{\sigma}_{b} \bar{\sigma}_{c} 
        \bar{\sigma}_{d} - h_{a} \bar{\sigma}_{a}
\ee
is the tree-level potential and $\bar{\sigma}_{a}$ 
is determined on the tree level by 
\be 
\frac{\partial U(\bar{\sigma})}{\partial 
        \bar{\sigma_{a}}}= m^2 \, \bar{\sigma}_{a} - 
        3\,{\cal G}_{abc} \bar{\sigma}_{b} 
        \bar{\sigma}_{c}+ \frac{4}{3} {\cal F}_{abcd} 
        \bar{\sigma}_{b} \bar{\sigma}_{c} 
        \bar{\sigma}_{d} - h_{a} = 0 \,\, .
\ee
The coefficients ${\cal G}_{abc}$, ${\cal F}_{abcd}$, and 
${\cal H}_{abcd}$ are given by
\begin{mathletters} 
\bea
{\cal G}_{abc} &=& \frac{c}{6} \left[ d_{abc} 
        - \frac{3}{2} \left(\delta_{a0} d_{0bc} +
        \delta_{b0} d_{a0c} + \delta_{c0} d_{ab0} \right) + 
        \frac{9}{2} d_{000} \delta_{a0} 
        \delta_{b0} \delta_{c0}\right] \,\, ,  \\
{\cal F}_{abcd} &=& \frac{\lambda_{1}}{4}  \, 
        \left(\delta_{ab} \delta_{cd} + 
        \delta_{ad} \delta_{bc} + \delta_{ac} \delta_{bd} \right) + 
        \frac{\lambda_{2}}{8} \, \left(d_{abn} d_{ncd} + 
        d_{adn} d_{nbc} + d_{acn} d_{nbd} \right) \,\, ,\\
{\cal H}_{abcd} &=& \frac{\lambda_{1}}{4}  \, \delta_{ab} \delta_{cd}  + 
        \frac{\lambda_{2}}{8} \, \left(d_{abn} d_{ncd} + 
        f_{acn} f_{nbd} + f_{bcn} f_{nad} \right) \,\,\, .
\eea
\end{mathletters}
The tree-level masses, $(m_{S}^{2})_{ab}$ and $(m_{P}^{2})_{ab}$
are given by
\begin{mathletters} \label{massmatrices}
\bea
(m_{S}^{2})_{ab} &=& m^{2} \,  \delta_{ab} - 6 \,  {\cal G}_{abc} \, 
         \bar{\sigma}_{c} 
        + 4  \, {\cal F}_{abcd} \,  
        \bar{\sigma}_{c} \bar{\sigma}_{d} \,\, , \label{massmatrixs}\\
(m_{P}^{2})_{ab} &=& m^{2} \,  \delta_{ab} + 6  \, {\cal G}_{abc} \, 
         \bar{\sigma}_{c} 
        + 4  \, {\cal H}_{abcd}  \, \bar{\sigma}_{c} 
        \bar{\sigma}_{d} \,\, .\label{massmatrixp}
\eea
\end{mathletters}
In general, these mass matrices are not diagonal. Consequently, 
the fields ($\sigma_{a}$, $\pi_{a}$) in the standard
basis of $U(3)$ generators are not mass eigenstates. 
Since the mass matrices are symmetric and real, diagonalization is 
achieved by an orthogonal transformation, 
\begin{mathletters}\label{ortho}
\bea 
\tilde{\sigma}_{i} &=& O^{(S)}_{ia} \, \sigma_{a} \,\, , \\ 
\tilde{\pi}_{i} &=& O^{(P)}_{ia} \, \pi_{a} \,\, , \\ 
\left(\tilde{m}_{S,P}^{2}\right)_{i} &=& O^{(S,P)}_{ai} \, 
        \left(m_{S,P}^{2}\right)_{ab} \, 
        O^{(S,P)}_{bi} \,\,. \label{orthoc}
\eea
\end{mathletters}

\section{Patterns of symmetry breaking} \label{III}

In this section, we discuss possible patterns of 
symmetry breaking in the vacuum.  We begin with the most symmetric 
case, i.e., with the minimum number of nontrivial 
couplings, and then successively reduce the symmetry.  
\begin{enumerate}
\item \underline{$H = 0$, $c = 0$, $\lambda_{2} = 0$:} 
For $m^{2} > 0$, 
the symmetry group is $O(18)$, on account of 
\be 
{\rm Tr} (\Phi^{\dagger} \Phi) = \frac{1}{2} (\sigma_{a}^{2} 
+ \pi_{a}^{2}) \,\, . 
\ee
The physics of the $O(N)$ 
model has been studied extensively in the 
past \cite{GellMann,JDRenorm}
and so we shall restrict ourselves in the 
following to $\lambda_{2} \neq 0$.
Here, we only mention that for $m^{2} < 0$, the $O(18)$ 
symmetry is spontaneously broken to $O(17)$ and there 
are 17 Goldstone bosons.

\item \underline{$H=0$, $c=0$, $\lambda_{2} \neq 0$:}
For $m^{2} > 0$, the Lagrangian has a global 
$SU(3)_{V} \times U(3)_{A}$ symmetry.  
For $m^{2} < 0$, Eq.\ (\ref{Utree}) shows that 
$\Phi$ develops a non-vanishing vacuum expectation 
value, and the $U(3)_{A}$ symmetry is {\em spontaneously\/} broken.
By the Vafa--Witten theorem \cite{Ed}, only the axial
symmetries can be spontaneously broken, while the vector
symmetries remain intact.
One can distinguish two cases \cite{Paterson}: 
\begin{enumerate}
\item $\lambda_{2} > 0$. 
$SU(3)_{V} \times U(3)_{A}$ is
broken to $SU(3)_{V}$, with $\langle \Phi \rangle 
\sim {\rm diag}(1,1,1)$  and 
the appearance of 9 Goldstone bosons which comprise the 
entire pseudoscalar nonet, i.e., the pions, the kaons, the $\eta$, and 
the $\eta'$ meson.  The nine massive scalar particles 
fall into irreducible representations of $SU(3)_{V}$. 
Since the mesons consist of a quark (a
$[{\bf 3}]$ of $SU(3)_{V}$)
and an antiquark (a $[{\bf \bar{3}}]$), these representations 
are a singlet and an octet,  because
$[{\bf 3}] \times [{\bf \bar{3}}] = 
[{\bf 1}] + [{\bf 8}]$.
The mass of the singlet, the $\sigma$ meson, 
is in general different from the (degenerate) 
masses of the octet particles.
\item $\lambda_{2} < 0$.
$SU(3)_{V} \times U(3)_{A}$
is broken to $SU(2)_{V} \times U(2)_{A}$, 
with $\langle \Phi \rangle \sim {\rm diag}(0,0,1)$ and 
10 Goldstone bosons.
\end{enumerate}

\item \underline{$H = 0$, $c \neq 0$, $\lambda_{2} \neq 0$:}
The symmetry is 
$SU(3)_{V} \times SU(3)_{A}$.
A non-vanishing $\langle \Phi \rangle$ spontaneously breaks 
this symmetry to $SU(3)_{V}$, with 
the appearance of 8 Goldstone bosons, which is 
the complete pseudoscalar octet.  The ninth Goldstone 
boson of case 2(a), the $\eta'$ meson, becomes massive and thus is
no longer a Goldstone boson, 
because the $U(1)_{A}$ 
symmetry is already explicitly broken.  
The masses of the scalar particles behave as in case 2(a).
Note that from Eq.\ (\ref{Utree}), 
$m^{2} < 0$ is no longer required for 
spontaneous symmetry breaking when $c \neq 0$. 

\item \underline{$H \neq 0$, $c = 0$, $\lambda_{2} \neq 0$:}
In QCD, this corresponds to non-vanishing 
quark masses, but vanishing $U(1)_{A}$ anomaly.  
Since $\langle \Phi \rangle$ must carry the quantum 
numbers of the vacuum, 
only $h_{0}$, $h_{3}$, and $h_{8}$ can be nonzero. 
One can distinguish three cases:

\begin{enumerate}

\item $h_{0} \neq 0$, $h_{3} = h_{8} = 0$.
All quark masses are equal, i.e., $m_{u} = m_{d} = m_{s}$.
In this case, the $U(3)_{A}$ axial symmetry is 
{\em explicitly\/} broken, i.e., the 9 Goldstone bosons of case 2(a) 
become (mass degenerate) pseudo-Goldstone bosons.  The 
$SU(3)_{V}$ symmetry 
remains intact and the scalars follow the classification as discussed 
in case 2(a). 

\item $h_{0} \neq 0$, $h_{3} = 0$, $h_{8} \neq 0$.
Only the non-strange flavors are degenerate in mass, i.e.,
$m_{u} = m_{d} \neq m_{s}$.  Besides the explicitly 
broken $U(3)_{A}$ symmetry, 
$SU(3)_{V}$ is explicitly broken to $SU(2)_{V}$.
For the scalar particles, the following applies.
If there was ideal flavor mixing, i.e., the 
physical particles are also eigenstates of flavor, one 
particle is an $s \bar{s}$ state (the $f_{0}$ meson), while all others 
contain at least one non-strange quark or antiquark.  
The latter then fall into   
irreducible representations of $SU(2)_{V}$. 
For the scalar particles containing no strange quark 
or antiquark, 
a quark (a $[ {\bf 2}]$ of $SU(2)_{V}$) couples with an 
antiquark (a $[ {\bf \bar{2}}]$) to form 
the $\sigma$ meson singlet and the $a_{0}$ meson triplet, since 
$[{\bf 2}] \times [{\bf \bar{2}}] = 
[{\bf 1}] + [{\bf 3}]$.
The strange scalar particles have only one 
quark in a $[{\bf 2}]$ or a $[{\bf \bar{2}}]$ representation, 
and therefore fall into doublets of $SU(2)_{V}$. 
The $\kappa^{+}$ and $\kappa^{0}$ mesons form a $[{\bf 2}]$, 
while the $\kappa^{-}$ and $\bar{\kappa}^{0}$ mesons form 
a $[{\bf \bar{2}}]$.
Because the masses of quarks and antiquarks are 
identical, these two doublets are mass degenerate.
In nature, however, flavor mixing is not ideal and 
the $f_0$ meson has a $q \bar{q}$ admixture, just as the 
$\sigma$ meson has an $s \bar{s}$ admixture.
For the pseudoscalars, only the four non-strange 
pseudo-Goldstone bosons, the pions and the $\eta'$ meson, 
are degenerate in mass.
The kaons and the $\eta$ meson both have different masses. 
Since the pions are non-strange $q \bar{q}$ states, and
the $\eta'$ meson is degenerate in mass with the pions, it follows that
it is also a non-strange $q \bar{q}$ state. Then, the $\eta$ meson is a
pure $s \bar{s}$ state,  i.e., flavor mixing is ideal
in the pseudoscalar sector.

\item $h_{0} \neq 0$, $h_{3} \neq 0$, $h_{8} \neq 0$.
Here, $SU(3)_{V}$ is completely broken. 
Even the non-strange pseudo-Goldstone bosons are no longer
completely degenerate in mass. 

\end{enumerate}

\item \underline{$H \neq 0$, $c \neq 0$, $\lambda_{2} \neq 0$:}
Now, from the $U(3)_{A} \cong SU(3)_{A} \times U(1)_{A}$ symmetry, 
the $U(1)_{A}$ is explicitly broken by instantons. 
Again, there are three cases: 

\begin{enumerate}

\item $h_{0} \neq 0$, $h_{3} = h_{8} = 0$.
In this case, the remaining $SU(3)_{A}$ axial symmetry is 
explicitly broken, i.e., the 8 Goldstone bosons of case 3
become (mass degenerate) pseudo-Goldstone bosons. As above, the 
$SU(3)_{V}$ symmetry 
remains intact.  The scalar particles behave as in case 4(a).    

\item $h_{0} \neq 0$, $h_{3} = 0$, $h_{8} \neq 0$.
Besides the explicitly 
broken $SU(3)_{A}$ symmetry, 
$SU(3)_{V}$ is explicitly broken to 
$SU(2)_{V}$.  
The scalar and pseudoscalar particles behave as in case 4(b),
except that the $\eta'$ meson mass is different from the pion mass
because of the $U(1)_{A}$ anomaly. Then, flavor
mixing is no longer ideal in the pseudoscalar sector.

\item $h_{0} \neq 0$, $h_{3} \neq 0$, $h_{8} \neq 0$.
This is the case realized in nature, although 
violation of isospin $SU(2)_{V}$ is small (the charged and 
the neutral pions are almost degenerate in mass).
\end{enumerate}

\end{enumerate}

In the following, we shall restrict ourselves to 
studying cases 2(a), 3, 4(b), and 5(b). 
The first two cases are interesting because they 
represent the idealized scenario where the quark 
masses are zero, i.e., the chiral limit.  
The last two cases are close to the situation in 
nature where quark masses break the chiral symmetry 
explicitly. Lattice QCD data 
indicate that the $U(1)_{A}$ anomaly becomes 
small for large temperatures \cite{latticeu1a}.  This 
motivates our interest in cases 2(a) and 4(b), see also 
\cite{Jurgen1}.
Since isospin $SU(2)_{V}$ violation is rather small 
in nature, it is sufficient and easier to study case 
5(b) instead of 5(c).  

\section{Condensates and masses in the vacuum} \label{IV}

In this section, we establish the vacuum properties of 
the $SU(3)_{r} \times SU(3)_{\ell}$ model in the 
various cases selected above. In 
contrast to the previous section, it is now easier 
to begin with the most asymmetric case 5(b).  This is 
the case closest to nature and therefore it is 
natural to use the experimental values for 
the meson masses and decay constants as input 
to determine the coupling constants. 
The linear sigma model has six parameters, $m^2$, $\lambda_1$, 
$\lambda_2$, $c$, $h_0$, and $h_8$. It therefore requires
six experimentally known quantities as input. We choose
$m_\pi$, $m_K$, $f_\pi$, $f_K$, the average squared mass of
the $\eta$ and $\eta'$ mesons, $m_{\eta}^2 + m_{\eta'}^2$, and
$m_\sigma$. The other masses, $m_{a_0}$, $m_\kappa$, $m_{f_0}$, 
the difference $m_{\eta}^2 - m_{\eta'}^2$, and the mixing angles
are then predicted.

In the other cases, nature does not provide us with 
information about the masses and the decay constants. 
In case 4(b), $c=0$, and we need to specify only five input
parameters. It turns out that we do not need
to fix $m_{\eta}^2 + m_{\eta'}^2$, because $m_{\eta'}= m_\pi$,
and the mass of the $\eta$ meson is predicted as well.

The cases 2(a) and 3 correspond to the chiral limit which is
not realized in nature. Therefore, in principle one should not use 
experimental values for the masses and decay constants to fix the
parameters $m^2$, $\lambda_1$, $\lambda_2$, and, in case 3, $c$.
For the sake of definiteness, however, we 
use the following tentative generalizations of experimental data.
We use the pion decay constant, $f_\pi$, extrapolated to the chiral limit,
the $\sigma$ meson mass, and, in case 3, the $\eta'$ meson mass.
[In case 2(a), $m_{\eta'} = m_\pi = 0$.]
For the final input parameter,
since the scalar octet is degenerate in mass when the $SU(3)_r \times
SU(3)_\ell$ symmetry is not explicitly broken, we use an average of
the experimental values for the masses of the scalar octet.

\subsection{Explicit chiral symmetry breaking with $U(1)_A$ anomaly}
\label{IVa}

The vacuum expectation value is $\langle \Phi \rangle
= T_0 \, \bar{\sigma}_0 + T_8 \, \bar{\sigma}_8$.
The equations for the condensates $\bar{\sigma}_0$ and $\bar{\sigma}_8$
read
\begin{mathletters} \label{heqs}
\bea
h_0 & = & \left[m^{2} - \frac{c}{\sqrt{6}}\, \bar{\sigma}_0
        + \left( \lambda_{1} + \frac{\lambda_{2}}{3} 
        \right) \bar{\sigma}_{0}^{2} \right] \, \bar{\sigma}_{0}
        + \left[ \frac{c}{2 \sqrt{6}} + \left( \lambda_1
        + \lambda_2 \right) \bar{\sigma}_0 - \frac{\lambda_2}{3 \sqrt{2}}
        \, \bar{\sigma}_8 \right] \, \bar{\sigma}_8^2 \,\, ,\label{h0} \\
h_8 & = & \left[m^{2} + \frac{c}{\sqrt{6}}\, \bar{\sigma}_0
        + \frac{c}{2 \sqrt{3}} \, \bar{\sigma}_8
        + \left( \lambda_{1} + \lambda_{2}\right) \bar{\sigma}_{0}^{2}
        - \frac{\lambda_2}{\sqrt{2}}\, \bar{\sigma}_0 \, \bar{\sigma}_8
        + \left( \lambda_1 +\frac{\lambda_2}{2}\,\right) 
        \bar{\sigma}_8^2 \right] \, \bar{\sigma}_8 \,\,  . \label{h8}
\eea
\end{mathletters}
The PCAC relations (see Appendix) determine the values of the
condensates from the pion and kaon decay constants, $f_\pi$, $f_K$,
\begin{mathletters} \label{SIGMA_08}
\bea
\bar{\sigma}_0 & = & \frac{f_\pi + 2\, f_K}{\sqrt{6}} \, \, , \\
\bar{\sigma}_8 & = & \frac{2}{\sqrt{3}} \left( f_\pi - f_K \right) \, \, .
\eea
\end{mathletters}
We use the experimental values $f_\pi = 92.4$ MeV, $f_K = 113$ MeV
\cite{PDB}.

The nonzero elements of the scalar mass matrix are
\begin{mathletters} \label{smass}
\bea
\left( m_S^2 \right)_{00} & = & m^2 - \sqrt{\frac{2}{3}}\, c \,
        \bar{\sigma}_0 + \left( 3\lambda_1 + \lambda_2 \right)
        \bar{\sigma}_0^2 + \left( \lambda_1 + \lambda_2 \right)
        \bar{\sigma}_8^2 \,\, , \\
\left( m_S^2 \right)_{11} & = & \left( m_S^2 \right)_{22} =
        \left( m_S^2 \right)_{33} \nonumber \\
        & = & m^2 + \frac{c}{\sqrt{6}} \,
        \bar{\sigma}_0 - \frac{c}{\sqrt{3}}\, \bar{\sigma}_8
        + \left( \lambda_1 + \lambda_2 \right)\bar{\sigma}_0^2 
        + \sqrt{2}\, \lambda_2 \, \bar{\sigma}_0 \, \bar{\sigma}_8
        + \left( \lambda_1 + \frac{\lambda_2}{2} \right)
        \bar{\sigma}_8^2 \,\, , \\
\left( m_S^2 \right)_{44} & = & \left( m_S^2 \right)_{55} =
        \left( m_S^2 \right)_{66} =\left( m_S^2 \right)_{77} 
        \nonumber \\
        & = & m^2 + \frac{c}{\sqrt{6}} \,
        \bar{\sigma}_0 + \frac{c}{2\sqrt{3}}\, \bar{\sigma}_8
        + \left( \lambda_1 + \lambda_2 \right)\bar{\sigma}_0^2 
        - \frac{ \lambda_2 }{\sqrt{2}}\, \bar{\sigma}_0 \, 
        \bar{\sigma}_8
        + \left( \lambda_1 + \frac{\lambda_2}{2} \right)
        \bar{\sigma}_8^2 \,\, , \label{mkappa}\\
\left( m_S^2 \right)_{88} & = & m^2 + \frac{c}{\sqrt{6}} \,
        \bar{\sigma}_0 + \frac{c}{\sqrt{3}}\, \bar{\sigma}_8
        + \left( \lambda_1 + \lambda_2 \right)\bar{\sigma}_0^2 
        - \sqrt{2}\, \lambda_2 \, \bar{\sigma}_0 \, 
        \bar{\sigma}_8
        + 3 \left( \lambda_1 + \frac{\lambda_2}{2} \right)
        \bar{\sigma}_8^2 \,\, , \\
\left( m_S^2 \right)_{08} & = & \left( m_S^2 \right)_{80} 
         =  \left[\frac{c}{\sqrt{6}} 
        + 2 \left( \lambda_1 + \lambda_2 \right)\bar{\sigma}_0
        - \frac{\lambda_2}{\sqrt{2}}\,  \bar{\sigma}_8
        \right]\, \bar{\sigma}_8 \,\, .
\eea
\end{mathletters}
While the masses of the $a_0$ and the $\kappa$ mesons are given by
the (11) and (44) elements of the mass matrix,
$m_{a_0}^2 \equiv (m_S^2)_{11}$, $m_{\kappa}^2 \equiv (m_S^2)_{44}$,
the $\sigma$ and $f_0$ meson masses are obtained by diagonalizing
the (08) sector of the mass matrix. According to Eq.\ (\ref{ortho}),
\begin{mathletters}
\bea
m_\sigma^2 & \equiv & (\tilde{m}_S^2)_0 = 
        (m_S^{2})_{00}\, \cos^{2}\theta_{S} + 
        (m_S^{2})_{88}\, \sin^{2}\theta_{S}  + 
        2\, (m_S^{2})_{08}\, \cos \theta_{S}\, \sin \theta_{S} \,\, , \\
m_{f_0}^2 & \equiv & (\tilde{m}_S^2)_8 = 
        (m_S^{2})_{00}\, \sin^{2}\theta_{S}  + 
         (m_S^{2})_{88}\, \cos^{2}\theta_{S} -
        2\, (m_S^{2})_{08}\, \cos \theta_{S}\, \sin \theta_{S} \,\, , 
\eea
where the scalar mixing angle $\theta_S$ is given by
\be
\tan 2 \theta_S  = \frac{2 \, (m_S^2)_{08}}{(m_S^{2})_{00} -
        (m_S^{2})_{88} } \,\, .
\ee
\end{mathletters}

The pseudoscalar mass matrix is given by 
\begin{mathletters} \label{psmass}
\bea
\left( m_P^2 \right)_{00} & = & m^2 + \sqrt{\frac{2}{3}}\, c \,
        \bar{\sigma}_0 + \left( \lambda_1 + \frac{\lambda_2}{3} \right)
        \left(\bar{\sigma}_0^2 + \bar{\sigma}_8^2 \right)\,\, , \\
\left( m_P^2 \right)_{11} & = & \left( m_P^2 \right)_{22} = 
        \left( m_P^2 \right)_{33} \nonumber \\
        & = & m^2 - \frac{c}{\sqrt{6}} \,
        \bar{\sigma}_0 + \frac{c}{\sqrt{3}}\, \bar{\sigma}_8
        + \left( \lambda_1 + \frac{\lambda_2}{3} \right)\bar{\sigma}_0^2 
        + \frac{\sqrt{2}}{3}\, \lambda_2 \, 
        \bar{\sigma}_0 \, \bar{\sigma}_8
        + \left( \lambda_1 + \frac{\lambda_2}{6} \right)
        \bar{\sigma}_8^2 \,\, , \label{mpi} \\
\left( m_P^2 \right)_{44} & = & \left( m_P^2 \right)_{55} =
        \left( m_P^2 \right)_{66} =\left( m_P^2 \right)_{77} 
        \nonumber \\
        & = & m^2 - \frac{c}{\sqrt{6}} \,
        \bar{\sigma}_0 - \frac{c}{2\sqrt{3}}\, \bar{\sigma}_8
        + \left( \lambda_1 + \frac{\lambda_2}{3} \right)\bar{\sigma}_0^2 
        - \frac{ \lambda_2 }{3\sqrt{2}}\, \bar{\sigma}_0 \, 
        \bar{\sigma}_8
        + \left( \lambda_1 + \frac{7}{6} \, \lambda_2 \right)
        \bar{\sigma}_8^2 \,\, , \label{mk} \\
\left( m_P^2 \right)_{88} & = & m^2 - \frac{c}{\sqrt{6}} \,
        \bar{\sigma}_0 - \frac{c}{\sqrt{3}}\, \bar{\sigma}_8
        + \left( \lambda_1 + \frac{\lambda_2}{3} \right)\bar{\sigma}_0^2 
        - \frac{\sqrt{2}}{3}\, \lambda_2 \, \bar{\sigma}_0 \, 
        \bar{\sigma}_8
        + \left( \lambda_1 + \frac{\lambda_2}{2} \right)
        \bar{\sigma}_8^2 \,\, , \\
\left( m_P^2 \right)_{08} & = & \left( m_P^2 \right)_{80} 
         =  \left[-\frac{c}{\sqrt{6}} 
        +  \frac{2}{3}\, \lambda_2 \, \bar{\sigma}_0
        - \frac{\lambda_2}{3\sqrt{2}}\, \bar{\sigma}_8
        \right]\, \bar{\sigma}_8 \,\, .
\eea
\end{mathletters}
While the mass of the pion and the kaon are given by
the (11) and (44) elements of the mass matrix,
$m_{\pi}^2 \equiv (m_P^2)_{11}$, $m_{K}^2 \equiv (m_P^2)_{44}$,
the $\eta'$ and $\eta$ meson masses are obtained by diagonalizing
the (08) sector of the mass matrix. According to Eq.\ (\ref{ortho}),
\begin{mathletters}
\bea
m_{\eta'}^2 & \equiv & (\tilde{m}_P^2)_0 =
        (m_P^{2})_{00}\, \cos^{2}\theta_{P}   + 
        (m_P^{2})_{88}\, \sin^{2}\theta_{P}  + 
        2\, (m_P^{2})_{08}\, \cos \theta_{P}\, \sin \theta_{P} \,\, , \\
m_{\eta}^2 & \equiv & (\tilde{m}_P^2)_8 = 
        (m_P^{2})_{00}\, \sin^{2}\theta_{P}   + 
         (m_P^{2})_{88}\, \cos^{2}\theta_{P} -
        2\, (m_P^{2})_{08}\, \cos \theta_{P}\, \sin \theta_{P} \,\, , 
\eea
where the pseudoscalar mixing angle $\theta_P$ is given by
\be
\tan 2 \theta_P  = \frac{2 \, (m_P^2)_{08}}{(m_P^{2})_{00} -
        (m_P^{2})_{88} } \,\, .
\ee
\end{mathletters}

The explicit symmetry breaking terms, $h_{0}$ and $h_{8}$, 
are determined from Eqs.\ (\ref{heqs}), (\ref{SIGMA_08}), (\ref{mpi}), 
and (\ref{mk}) as 
\begin{mathletters} \label{hdecay}
\bea
h_{0} &=& \frac{1}{\sqrt{6}} \left( m_{\pi}^{2} \, f_{\pi} + 
        2\, m_{K}^{2} \, f_{K} \right) \,\, , \\
h_{8} &=& \frac{2}{\sqrt{3}} \left( m_{\pi}^{2} \, f_{\pi} - 
        m_{K}^{2} \, f_{K} \right) \,\, .
\eea
\end{mathletters}
Using the experimental pion and kaon masses, one obtains
$h_0 = (286.094 \, {\rm MeV})^3$ and $h_8 = - (310.960\, {\rm MeV})^3$.

Comparing Eqs.\ (\ref{h8}) and (\ref{mkappa}) reveals 
\be \label{mkappa2}
h_{8} = m_{\kappa}^{2} \, \bar{\sigma}_{8} \,\, ,
\ee
i.e., the mass of the $\kappa$ meson is predicted to be 
$m_{\kappa} = 1124.315$ MeV, which is about 21\%
smaller than the experimental value. 
The average $\eta$ and $\eta'$ meson mass squared determines 
$\lambda_{2}$ 
\be
\lambda_{2} = \frac{3 \left( 2 \, f_{K} - 
        f_{\pi}\right) \, m_{K}^{2} - 
        \left( 2 \, f_{K} +f_{\pi}\right) \, m_{\pi}^2
        -2 \,  \left( m_{\eta'}^2 + m_{\eta}^2 \right) \, 
        \left( f_{K} - f_{\pi}\right)}
        {\left[3 \, f_{\pi}^2 + 8 \, f_{K} (f_{K} - 
        f_{\pi})\right] \, (f_{K} - f_{\pi})} \,\, .
\ee
For $m_{\eta'}^2 + m_{\eta}^2 = (1103.625\, {\rm MeV})^{2}$, 
one obtains $\lambda_{2} = 46.484$.  

The difference of the pion and kaon masses squared and 
$\lambda_2$ determine $c$, 
\be
c = \frac{m_{K}^{2} - m_{\pi}^2}{f_{K} - f_{\pi}} - \lambda_2 \,
        \left(2 \, f_{K} - f_{\pi} \right) = 4807.835\, {\rm MeV}\,\, .
\ee
Now, also the mass of the $a_{0}$ meson is fixed, 
\be
m_{a_{0}}^{2} = m_{\kappa}^{2} + (f_{K} - f_{\pi}) \left[c - 
        \lambda_2 \, (2\, f_{K} + f_{\pi}) \right] = (1028.707 
        \,{\rm MeV})^{2} \,\, ,
\ee
which is about 4\% larger than the experimental value. 
The pseudoscalar mixing angle is $\theta_{P} = -5^{{\rm o}}$.
This angle determines the individual $\eta$ and $\eta'$ meson masses 
to be $m_{\eta} = 539.008$ MeV and $m_{\eta'} = 963.046$ MeV.
These values are surprisingly 
close to the experimental ones; the $\eta$ meson is about
2\% lighter and the $\eta'$ meson is about 0.6\% heavier than in 
nature.

Finally, $\lambda_1$ is determined by fixing either the 
mass of the $\sigma$ or the $f_{0}$ meson; 
the other mass, the scalar mixing angle, $\theta_{S}$, and 
$m^{2}$ are then given by solving a nonlinear equation 
for the fixed mass.  
Here, we choose $m_{\sigma} = 600$ MeV to yield $\lambda_1 = 1.400$ and
$m_{f_{0}} = 1221.113$ MeV, about 11\% smaller 
than the experimental value.  The 
scalar mixing angle is $\theta_{S} = 19.859^{{\rm o}}$, and 
$m^{2} = (342.523 \, {\rm MeV})^{2}$. 
 
\subsection{Explicit chiral symmetry breaking without
$U(1)_A$ anomaly}

\label{IVb}

In the absence of the $U(1)_{A}$ anomaly, $c=0$, the 
condensate equations (\ref{h0}) and (\ref{h8}) and the 
equations for the scalar and pseudoscalar mass matrices 
(\ref{smass}) and (\ref{psmass}) simplify.  Equations 
(\ref{SIGMA_08}) and (\ref{hdecay}), however, remain the same 
and thus yield the same values for $\bar{\sigma}_{0}$, 
$\bar{\sigma}_{8}$, $h_0$, and $h_8$ as above.
The mass of the $\kappa$ meson is still given by 
Eq.\ (\ref{mkappa2}) and is the same as above.

The differences begin with the mass of the $\eta'$ meson, which 
is now identical to the pion mass, since the $U(1)_{A}$ anomaly
is absent. Furthermore,
$\lambda_2$ is given by the kaon and pion masses and decay constants,
\be
\lambda_2 = \frac{m_K^2 - m_\pi^2}{(2\, f_K - f_\pi) (f_K-f_\pi)}
= 82.470 \,\, .
\ee
This then determines the mass of the $\eta$ meson,
\be
m_\eta^2 = m_\pi^2 + 2\, \lambda_2 \, f_K \, (f_K - f_\pi) = 
(634.818\, {\rm MeV})^2 \,\, ,
\ee
and the $a_0$ meson,
\be
m_{a_0}^2 = m_\kappa^2 - \lambda_2 \left(2\, f_K^2 - f_\pi\, f_K
        - f_\pi^2\right) = (850.387 \, {\rm MeV})^2 \,\, .
\ee
The pseudoscalar mixing angle is given by
$\tan 2 \theta_P = 2 \sqrt{2}$, or $\theta_P = 35.264^{\rm o}$.
This corresponds to ideal flavor mixing.

As before, $\lambda_1$ is given by solving the equation for
the $\sigma$ meson mass, which in turn yields the $f_0$ meson mass,
the scalar mixing angle, and $m^2$. For $m_\sigma = 600$ MeV we find
$\lambda_1 = - 4.550$, $m_{f_0} = 1341.367$ MeV,
$\theta_S = 31.326^{\rm o}$, and $m^2 = -(503.551
\, {\rm MeV}^2)$.

\subsection{Chiral limit with $U(1)_A$ anomaly} \label{IVc}

In this case, $\langle \Phi \rangle = T_{0} \, \bar{\sigma}_{0}$. 
Using $h_{a} =  0$ and the 
explicit form for the ${\cal G}_{abc}$ and 
${\cal F}_{abcd}$, the equation 
for the condensate at the tree level (\ref{h0}) simplifies to
\be \label{3cond}
0 = \left[m^{2} - \frac{c}{\sqrt{6}}\, \bar{\sigma}_0
        + \left( \lambda_{1} + \frac{\lambda_{2}}{3} 
        \right) \bar{\sigma}_{0}^{2} \right] \, \bar{\sigma}_{0} \,\, .
\ee 
{}From the PCAC relations (see Appendix), we now obtain 
$\bar{\sigma}_{0} = \sqrt{3/2} \, f_{\pi}$. Lattice QCD data
indicate a linear behavior of $f_\pi$ with the quark mass, 
$f_\pi \simeq a\, m_q + b$ \cite{pcomblum}.
Extrapolating these data to the chiral limit, $m_q \rightarrow 0$
(as appropriate for $H=0$), one obtains $f_\pi = 90$ MeV.

The scalar and the pseudoscalar mass matrices are diagonal and have
the particularly simple form
\begin{mathletters}
\bea
\left(m_{S}^{2}\right)_{ab} &=& \left[ m^{2} 
        + \frac{c}{\sqrt{6}} \, \bar{\sigma}_0 +
        \left( \lambda_{1} + \lambda_{2} \right) \bar{\sigma}_{0}^{2}
        \right] \, \delta_{ab} - \left( \sqrt{\frac{3}{2}} \, c 
        - 2 \lambda_{1} \bar{\sigma}_{0} \right)
        \bar{\sigma}_0 \,\delta_{a0} \delta_{b0} \,\, , \\ 
\left(m_{P}^{2}\right)_{ab} &=& \left[m^{2} 
        - \frac{c}{\sqrt{6}} \, \bar{\sigma}_0 +
        \left( \lambda_{1} + \frac{\lambda_{2}}{3} 
        \right) \bar{\sigma}_{0}^{2} \right]
        \delta_{ab} + \sqrt{\frac{3}{2}}\, c \,
        \bar{\sigma}_0 \,\delta_{a0} \delta_{b0}  \,\, . \label{30b}
\eea
\end{mathletters}
On account of Eq.\ (\ref{3cond}), the pseudoscalar octet, 
$a,b=1,\ldots,8$, is massless, as expected from group theory, see
case 3 in Sec.\ \ref{III}. The singlet, the $\eta'$ meson, is massive,
because the $U(1)_{A}$ symmetry is explicitly broken by the $U(1)_A$ anomaly,
\be
m_{\eta'}^2 \equiv \left( m_P^2 \right)_{00}
        = \frac{3}{2} \, c \, f_\pi\,\, .
\ee
If we use the experimental value of the $\eta'$ meson mass, $c = 6791.157$ MeV.

The scalar particles fall into a singlet, the $\sigma$ meson,
with mass 
\be
m_{\sigma}^{2} \equiv \left(m_{S}^{2}\right)_{00} =  m^{2} 
        - c\, f_\pi + 
        \frac{3}{2} \left( 3 \, \lambda_{1} + 
        \lambda_{2} \right)  f_{\pi}^{2}\,\, , 
\ee
and an octet comprising the $a_{0}$, $\kappa$, and $f_{0}$ mesons, 
\be
m_{a_{0}}^{2} \equiv \left(m_{S}^{2}\right)_{11} = m^{2} 
        + \frac{c}{2}\, f_\pi + 
        \frac{3}{2} \left( \lambda_{1} + 
        \lambda_{2} \right)  f_{\pi}^{2} \equiv m_{\sigma}^{2}
        + m_{\eta'}^2 - 
        3 \lambda_{1} f_{\pi}^{2} \equiv \lambda_2 \, f_\pi^2
        + \frac{2}{3} \, m_{\eta'}^2\,\, ,
\ee 
where the last identity follows from Eq.\ (\ref{3cond}).
This equation determines $\lambda_{1}$ and $\lambda_2$ 
for given $m_{\sigma}$, $m_{a_{0}}$, and $m_{\eta'}$.  
As the octet is mass degenerate, we average the
experimental values for $m_{a_0}^2$, $m_{\kappa}^2$, and $m_{f_0}^2$,
weighted by the respective isospin degeneracy, 
to obtain an average mass of $m_{a_0} = 1225.795$ MeV.
For the $\sigma$ meson we again take $m_\sigma = 600$ MeV.
This results in $\lambda_1 = -9.291$ and
$\lambda_2 = 110.046$. Then, $m^2$ follows from
any of the mass equations as $m^2 = - (164.921\,{\rm MeV})^2$.

Note that the scalar singlet is lighter than the scalar octet,
but the pseudoscalar singlet is heavier than the pseudoscalar octet.
This ``inverted mass spectrum'' for the scalar mesons relative
to the pseudoscalar mesons \cite{Jaffe}
is a general feature in the presence of the $U(1)_A$ anomaly.
It arises from the relative difference
in sign of the terms $\sim {\cal G}_{abc}$ in Eqs.\ (\ref{massmatrices}).

\subsection{Chiral limit without $U(1)_A$ anomaly}
\label{IVd}

In this case, as the $U(1)_A$ symmetry is not explicitly broken,
all nine pseudoscalar particles are massless. This is readily
derived from Eqs.\ (\ref{3cond}) and (\ref{30b}) when $c=0$.

The scalar particles again fall into a singlet, the $\sigma$ meson,
with mass 
\be
m_{\sigma}^{2} \equiv \left(m_{S}^{2}\right)_{00} =  m^{2} + 
        \frac{3}{2} \left( 3 \, \lambda_{1} + 
        \lambda_{2} \right)  f_{\pi}^{2}\,\, , 
\ee
and an octet comprising the $a_{0}$, $\kappa$, and $f_{0}$ mesons, 
\be
m_{a_{0}}^{2} \equiv \left(m_{S}^{2}\right)_{11} = m^{2} + 
        \frac{3}{2} \left( \lambda_{1} + 
        \lambda_{2} \right)  f_{\pi}^{2} \equiv m_{\sigma}^{2} - 
        3 \lambda_{1} f_{\pi}^{2} \equiv \lambda_2 \, f_\pi^2\,\, ,
\ee 
where the last identity follows from Eq.\ (\ref{3cond}) with $c=0$.
This equation determines $\lambda_{1}$ and $\lambda_2$ 
for given $m_{\sigma}$ and 
$m_{a_{0}}$.  We again use $m_{a_0} = 1225.795$ MeV and
$m_\sigma = 600$ MeV.
This results in $\lambda_1 = -47.019$ and
$\lambda_2 = 185.503$. Then, $m^2$ follows from
any of the mass equations, $m^2 = - (424.264\,{\rm MeV})^2$, which
is negative, as required in this case for spontaneous symmetry breaking.

\section{The Effective Potential in the Cornwall--Jackiw--Tomboulis 
formalism} 
\label{V}

The effective potential of the  Cornwall--Jackiw--Tomboulis 
formalism \cite{CJT} is
\bea
V[\bar{\sigma},{\cal S},{\cal P}] &=& 
U(\bar{\sigma}) +\half \int_k \, 
\left\{ \left[\ln {\cal S}^{-1}(k)\right]_{aa} 
+ \left[\ln {\cal P}^{-1}(k)\right]_{aa}\right\} \nonumber \\
&+& \half\,\int_k\,   \left[ S^{-1}_{ab}(k;\bar{\sigma})
\, {\cal S}_{ba}(k) + P^{-1}_{ab}(k;\bar{\sigma})
\, {\cal P}_{ba}(k) - 2 \delta_{ab} \, \delta_{ba} \right] + 
V_2[\bar{\sigma},{\cal S},{\cal P}] \,\, . \label{V_CJT}
\eea
Here, $U(\bar{\sigma})$ is the tree-level potential of 
Eq.\ (\ref{Utree}), and 
\begin{mathletters}
\bea
S_{ab}^{-1}(k;\bar{\sigma}) &=& -k^2 \, \delta_{ab} + 
        \left(m_{S}^{2} \right)_{ab} \,\, , \\
P_{ab}^{-1}(k;\bar{\sigma}) &=& -k^2 \, \delta_{ab} + 
        \left(m_{P}^{2} \right)_{ab} \,\, , 
\eea
\end{mathletters}
are the tree-level propagators for scalar and 
pseudoscalar particles, with the respective mass matrices
(\ref{massmatrices}).  

The expectation values for the scalar fields, $\bar{\sigma}_{a}$, 
and the full propagators for scalar, ${\cal S}(k)$, and pseudoscalar, 
${\cal P}(k)$, particles are determined from the 
stationarity conditions
\begin{mathletters}
\bea \label{stationphi}
\frac{\delta V[\bar{\sigma},{\cal S},{\cal P}]}{\delta 
\bar{\sigma}_{a}} \, 
&=& 0 \,\, ,\\
 \frac{\delta V[\bar{\sigma},{\cal S},{\cal P}]}{\delta 
        {\cal S}_{ab}(k)} \,
&=& 0 \,\, , \label{stationS} \\
\frac{\delta V[\bar{\sigma},{\cal S},{\cal P}]}{\delta 
        {\cal P}_{ab}(k)} \,
&=& 0 \,\, . \label{stationP}
\eea
\end{mathletters}
With Eq.\ (\ref{V_CJT}), the latter two can be written in the form
\begin{mathletters} \label{schwinger}
\bea 
{\cal S}_{ab}^{-1}(k) &=& S_{ab}^{-1}(k;\bar{\sigma}) + 
        \Sigma_{ab}(k)\,\, , \\
{\cal P}_{ab}^{-1}(k) &=& P_{ab}^{-1}(k;\bar{\sigma}) + 
        \Pi_{ab}(k)\,\, , 
\eea
\end{mathletters}
where
\begin{mathletters}\label{selfenergy}
\bea
\Sigma_{ab}(k) &\equiv& 2 \, \frac{\delta V_2 [\bar{\sigma},
        {\cal S}, {\cal P}]}{\delta {\cal S}_{ba}(k)} \,\, , \\
\Pi_{ab}(k) &\equiv& 2 \, \frac{\delta V_2 [\bar{\sigma},
        {\cal S}, {\cal P}]}{\delta {\cal P}_{ba}(k)} \,\, , 
\eea
\end{mathletters}
are the self-energies for the scalar and pseudoscalar particles.

\vspace*{-1cm}
\begin{figure}
\hspace*{4.5cm}
\mbox{\epsfig{file=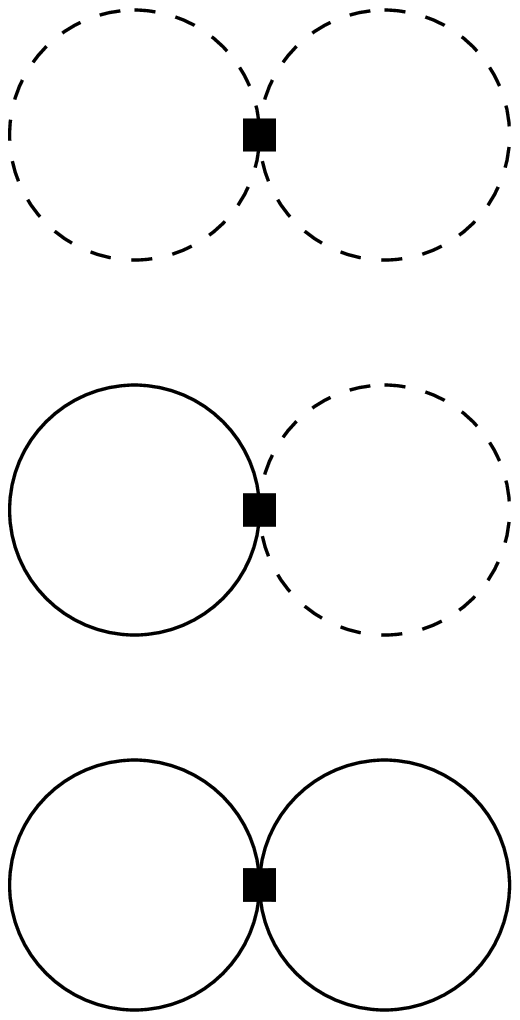,height=5cm,angle=270}}
\vspace{1cm}
\caption{The double-bubble diagrams.  Full lines 
are scalar particles, dashed lines are pseudoscalar 
particles.\label{fig1}}
\end{figure}

In general, $V_{2}[\bar{\sigma},{\cal S}, {\cal P}]$ is 
the sum of all two-particle irreducible (2PI) diagrams, with 
all lines representing full propagators.  Here, we 
restrict ourselves to the most simple class of 2PI diagrams, 
the double-bubble diagrams of Fig.\ \ref{fig1}, which 
is equivalent to the Hartree approximation.  
Explicitly, 
\be
V_{2}[\bar{\sigma},{\cal S}, {\cal P}] = {\cal F}_{abcd} \, \left[
        \int_{k} {\cal S}_{ab}(k) \int_{p} {\cal S}_{cd}(p) + 
        \int_{k} {\cal P}_{ab}(k) \int_{p} {\cal P}_{cd}(p) \right] +
        2 \, {\cal H}_{abcd} 
        \int_{k} {\cal S}_{ab}(k) \int_{p} {\cal P}_{cd}(p) \,\, .
\ee
Note that, in the Hartree approximation, $V_{2}$ is 
actually independent of $\bar{\sigma}_a$.  Therefore, 
the stationarity conditions for the condensates are 
\begin{mathletters}
\bea
h_{a} &=& m^{2} \, \bar{\sigma}_{a} - 3 \, {\cal G}_{abc} \,
        \left\{ \bar{\sigma}_{b} \, \bar{\sigma}_{c} + 
        \int_{k} \left[ {\cal S}_{cb}(k) - 
        {\cal P}_{cb}(k) \right] \right\} \nonumber \\ 
        &+& 4\, {\cal F}_{abcd} \, \left[ \frac{1}{3} \,
        \bar{\sigma}_{b} \, \bar{\sigma}_{c} + 
        \int_{k}  {\cal S}_{cb}(k) \right] \,  \bar{\sigma}_{d}
        + 4\, {\cal H}_{bcad} \,
        \bar{\sigma}_{d} \,  
        \int_{k}  {\cal P}_{cb}(k) \,\, . \label{Cond}
\eea
Since the self-energies (\ref{selfenergy}) are independent 
of momentum in the Hartree approximation, the 
Schwinger--Dyson equations (\ref{schwinger}) for the full propagators 
assume the simple form
\bea 
{\cal S}^{-1}_{ab}(k) &=& -k^2 \, \delta_{ab} + 
        \left( M_{S}^{2} \right)_{ab} \,\, , \label{invS}\\
{\cal P}^{-1}_{ab}(k) &=& -k^2 \, \delta_{ab} + 
        \left( M_{P}^{2} \right)_{ab} \,\, ,\label{invP}
\eea
\end{mathletters}
where the scalar and pseudoscalar mass matrices in the 
Hartree approximation are given by 
\begin{mathletters} \label{massmatrices2}
\bea
\left( M_{S}^{2} \right)_{ab} & = & m^{2} \, 
        \delta_{ab} - 6 \, {\cal G}_{abc} \,
        \bar{\sigma}_{c} + 
        4\, {\cal F}_{abcd} \, \left[ 
        \bar{\sigma}_{c} \, \bar{\sigma}_{d} + 
        \int_{k}  {\cal S}_{cd}(k) \right] 
        + 4\, {\cal H}_{abcd} \,
        \int_{k}  {\cal P}_{cd}(k) \,\, , \\
\left( M_{P}^{2} \right)_{ab} & = & m^{2} \, 
        \delta_{ab} + 6 \, {\cal G}_{abc} \,
        \bar{\sigma}_{c} + 
        4\, {\cal H}_{abcd} \, \left[ 
        \bar{\sigma}_{c} \, \bar{\sigma}_{d} + 
        \int_{k}  {\cal S}_{cd}(k) \right] 
        + 4\, {\cal F}_{abcd} \,
        \int_{k}  {\cal P}_{cd}(k) \,\, .
\eea
\end{mathletters}
In general, the mass matrices are not diagonal in the 
standard basis of $U(3)$ generators, see Sec.\ \ref{IV}.
Consequently, the propagators are also not diagonal in 
this basis.  Physically, however, only mass eigenstates can
propagate.  Therefore, we have to diagonalize the 
propagators before we compute the loop integrals in 
Eqs.\ (\ref{Cond}) and (\ref{massmatrices2}).

In the Hartree approximation, all particles are 
stable quasiparticles, i.e., the imaginary parts of the 
self-energies vanish.  Therefore, the inverse propagators 
(\ref{invS}) and (\ref{invP})
are real-valued.  They are also symmetric in the standard 
basis of $U(3)$ generators and thus diagonalizable via 
an orthogonal transformation.  This transformation is 
given by Eq.\ (\ref{orthoc}), with the obvious replacements
\be
\left( m_{S,P}^{2} \right)_{ab} \rightarrow 
\left( M_{S,P}^{2} \right)_{ab} \,\,\,\, , \,\,\,\, 
\left( \tilde{m}_{S,P}^{2} \right)_{i} \rightarrow 
\left( \tilde{M}_{S,P}^{2} \right)_{i}\,\, . \label{replace}
\ee

In cases 2(a) and 3 in Sec.\ \ref{III}, the mass matrices 
are diagonal at zero temperature, cf.\ Sec.\ \ref{IV}, and 
are taken to be diagonal at nonzero temperature $T$
as well.  In cases 4(b) 
and 5(b), they have off-diagonal elements in the (08) sector 
at zero temperature, and consequently also have off-diagonal 
contributions in 
this sector at nonzero temperature.  In the latter case, 
diagonalization proceeds as in Eq.\ (\ref{orthoc}) with the 
replacements (\ref{replace}).

The propagator matrices are diagonalized by the 
same orthogonal transformation as their inverse. 
The loop integrals in Eqs.\  (\ref{Cond}) and (\ref{massmatrices2})
are therefore computed, for example, as 
\be \label{rotS}
\int_k {\cal S}_{ab}(k) = O^{(S)}_{ai}
        \int_k \tilde{{\cal S}}_{i}(k) \,\, \,  O^{(S)}_{bi} \,\, , 
\ee
where $\tilde{{\cal S}}_{i}(k)$ is the scalar propagator 
in the mass eigenbasis. 

The loop integral in Eq.\ (\ref{rotS}) requires renormalization. 
Renormalization of many-body approximation schemes
is nontrivial \cite{JDRenorm}, but does not 
change the results qualitatively.  We therefore 
simply omit the vacuum contributions to the loop integrals and 
set 
\be
\int_k \tilde{{\cal S}}_{i}(k) = 
\int \frac{d^3{\bf k}}{(2 \pi)^3}\,  
    \frac{1}{\epsilon_{\bf k}[(\tilde{M}^{2}_{S})_{i}]} \, 
    \left(\exp\left\{\frac{\epsilon_{\bf k}[(\tilde{M}^{2}_{S})_{i}
        ]}{T}\right\}-1 \right)^{-1} \,\, ,
\ee
and similarly for the pseudoscalar loop integrals.  Here, 
$\epsilon_{\bf k}[(\tilde{M}^{2}_{S})_{i}] = 
\sqrt{{\bf k}^{2} + 
(\tilde{M}^{2}_{S})_{i}}$ is the relativistic energy of the $i$th 
scalar quasiparticle with momentum ${\bf k}$.

\section{Results} \label{VI}

In this section, we discuss the numerical results at 
nonzero temperature for the four cases of interest.

\subsection{Chiral limit}

\begin{figure}
\hspace*{1.cm}
\mbox{\epsfig{file=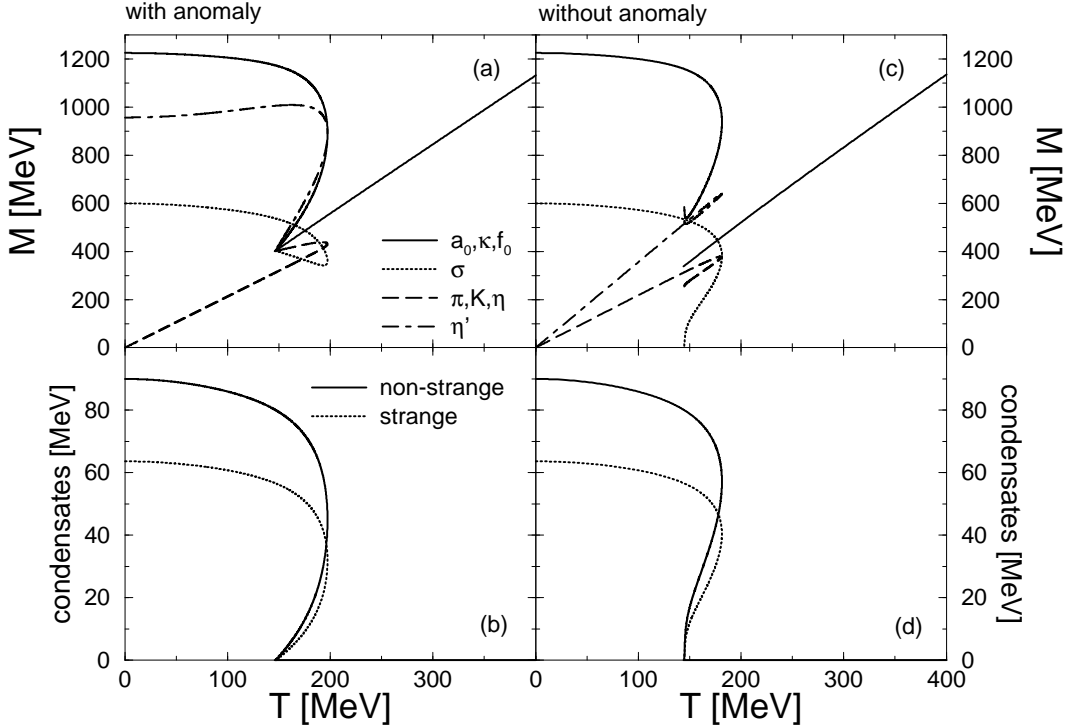,height=8cm,angle=270}}
\vspace{-1cm}
\caption{
The meson masses and the condensates as a function of 
temperature for (a,b) the case with $U(1)_A$ anomaly ($c \neq 0$)
and (c,d) the case without $U(1)_A$ anomaly ($c = 0$),
for $H=0$.
In (a,c), the full lines are the masses of the $a_{0}$, $\kappa$,
and $f_0$ mesons, the dotted lines represent the $\sigma$ meson 
mass, the dashed lines are $\pi$, $K$, and $\eta$ meson masses, 
and the dot-dashed lines are the $\eta'$ meson mass. 
In (b,d), the full lines represent the non-strange 
condensate, and the dotted lines represent the strange condensate.
\label{fig2}}
\end{figure}

In Fig.\ \ref{fig2}(a), the masses
are shown in the chiral limit with the $U(1)_{A}$ anomaly, 
$c\neq 0$.  This corresponds to case 3 of Sec.\ \ref{III},
for which the zero-temperature properties were 
discussed in Sec.\ \ref{IVc}. 
Accordingly, there are eight Goldstone bosons, the three 
pions, the four kaons and the $\eta$ meson, while the $\eta'$ meson has 
a large mass due to the $U(1)_{A}$ anomaly.  The scalar octet, 
comprising the three $a_0$ mesons, the four $\kappa$ mesons, 
and the $f_0$ meson, is mass degenerate, while the 
singlet, the $\sigma$ meson, has a different mass.  

As the temperature increases, the scalar masses decrease while 
the pseudoscalar masses increase.  The mass of the 
Goldstone bosons (the pseudoscalar octet) increases 
because the Hartree approximation does not 
respect Goldstone's theorem at nonzero $T$ \cite{JDRenorm}.
We do not consider this to be a serious shortcoming: 
as shown in \cite{Wambach}, supplementing the Hartree approximation 
with the Random Phase Approximation (RPA) cures this 
problem.  The analogous treatment within the 
CJT formalism will be deferred to a forthcoming 
publication \cite{lenrisforth}.

At some critical temperature, $T_{c} \sim 170$ MeV, there is a 
first order phase transition between the low-temperature
phase where chiral symmetry is broken and the high-temperature
phase where chiral symmetry is restored and all meson 
masses are equal.  The exact numerical value of $T_{c}$
can be determined by computing the effective potential. 

At the point where the condensates vanish, all masses 
become degenerate.  The reason is that at this point
the condensate equation enforces the constraint 
\be \label{constrained}
0 = \frac{c}{6 \sqrt{6}} \left\{ 2 \, \int_{k} 
        \left[ {\cal S}_{00}(k) - {\cal P}_{00}(k)
        \right] - \sum_{a=1}^{8} \int_{k} \left[ 
        {\cal S}_{aa}(k) - {\cal P}_{aa}(k) \right] \right\} \,\, ,
\ee
which is fulfilled when all masses are equal.   

In Fig.\ \ref{fig2}(b), the non-strange and strange 
condensates, $\varphi_{\rm ns}$ and $\varphi_{\rm s}$,
respectively, are shown as a function of temperature. 
In the standard basis of $U(3)$ generators, 
these are defined as 
\be
\langle \Phi \rangle  = \frac{1}{2} \left( \begin{array}{ccc}
        \varphi_{\rm ns} & 0 & 0 \\
        0 & \varphi_{\rm ns} & 0 \\  
        0 & 0 & \sqrt{2} \, \varphi_{\rm s} \end{array} \right) \,\, .
\ee
When $\langle \Phi \rangle = T_0 \, \bar{\sigma}_{0}$, 
$\varphi_{\rm ns} = \sqrt{2/3} \, \bar{\sigma}_{0}$, 
and  $\varphi_{\rm s} = \bar{\sigma}_{0} / \sqrt{3}$, 
i.e., $\varphi_{\rm s} = \varphi_{\rm ns}/\sqrt{2}$, 
as borne out by Fig.\ \ref{fig2}(b).

In Figs.\ \ref{fig2}(c,d), the masses and condensates 
are shown for the case without explicit $U(1)_{A}$ symmetry breaking, 
$c = 0$.  This corresponds to case 2(a) of Sec.\ \ref{III}. 
The zero-temperature properties were discussed in Sec.\
\ref{IVd}. 
Now, there are nine Goldstone bosons, the three 
pions, the four kaons, the $\eta$, and the $\eta'$ meson.
The behavior of the scalar octet is similar to the 
previous case.  

As the temperature increases, the behavior of the scalar 
and pseudoscalar masses is quite similar to the masses in 
Fig.\ \ref{fig2}(a), i.e., the scalar masses decrease and 
the pseudoscalar masses increase, until 
they become degenerate in a first order phase 
transition.  The critical temperature for this 
transition, however, appears to be slightly lower
than for $c \neq 0$.
A notable difference between Figs.\ \ref{fig2}(a)
and (c) is that the masses do not 
become degenerate continuously as the condensates vanish.  The reason is
that, for $c=0$, the above constraint 
equation (\ref{constrained}) is absent. However,
these phenomena are irrelevant for the thermodynamic properties
of the model, as they occur in a region where the solutions are
thermodynamically unstable and the stable solution has to be found from
Maxwell's construction for first order phase transitions.
\vspace*{0.5cm}

Another interesting feature is that the mass of the 
$\sigma$ meson is proportional to the condensates.
The reason for this is that the condensate equation 
and the equation for the $\sigma$ mass can 
be combined to give
\be
M^{2}_{\sigma} = \frac{8}{3} \, {\cal F}_{0000} \, \bar{\sigma}_{0}^{2} 
        \,\, .
\ee

\subsection{Explicit chiral symmetry breaking with $U(1)_A$ anomaly}

\begin{figure}
\hspace*{1cm}
\mbox{\epsfig{file=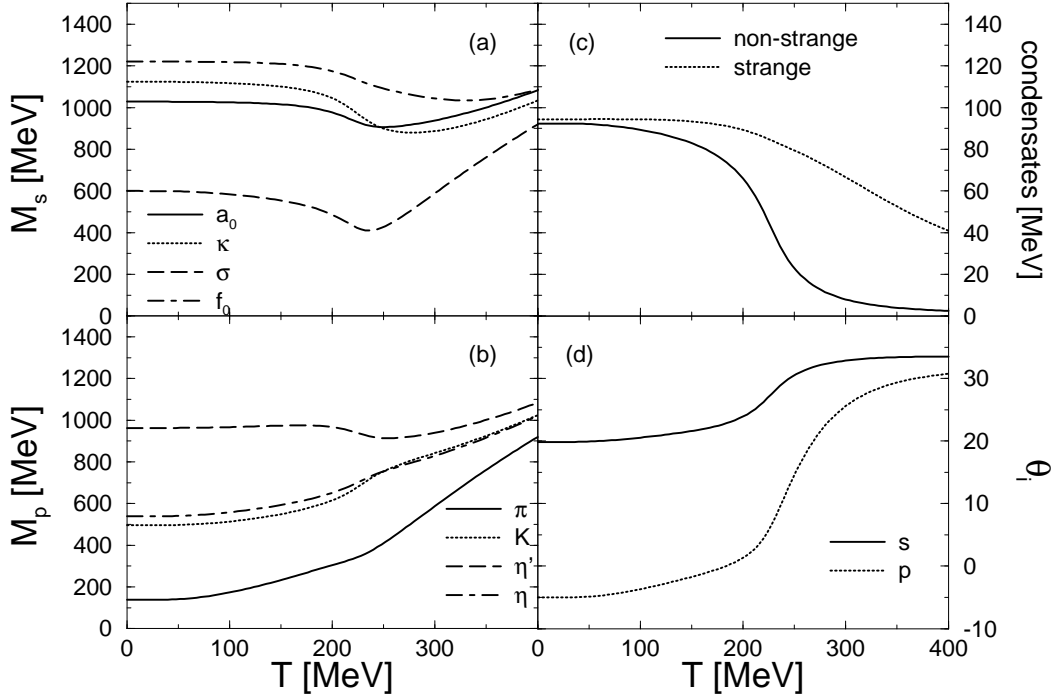,height=8cm,angle=270}}
\vspace{0cm}
\caption{The meson masses, the condensates, and 
the mixing angles as a function of 
temperature for $h_0,\, h_8 \neq 0$, $h_3=0$, and $c \neq 0$.
(a) For the scalar mesons, 
the full line is the $a_0$, the dotted line is the 
$\kappa$, the dashed line is the $\sigma$, and the 
dot-dashed line is the $f_{0}$ meson mass.  
(b) For the pseudoscalar mesons, 
the full line is the pion, the dotted line is the 
kaon, the dashed line is the $\eta'$, and the 
dot-dashed line is the $\eta$ meson mass. 
(c) The non-strange (full) and strange (dotted) condensates. 
(d) The scalar (full) and the pseudoscalar (dotted) 
mixing angles. 
\label{fig3}}
\end{figure}

In Fig.\ \ref{fig3}, we show the masses for the scalars (a), 
the pseudoscalars (b), the condensates (c), and the mixing 
angles (d) for explicit chiral symmetry breaking, including the
$U(1)_{A}$ anomaly. The masses behave according to the 
discussion of case 5(b) in Sec.\ \ref{III}.  As the temperature 
increases, chiral symmetry is restored in a 
crossover transition and all masses become approximately degenerate.
The temperature range of the crossover transition is $\sim 220$ MeV, i.e., 
about $50$ MeV higher than in the case where there is no 
explicit chiral symmetry breaking.  
A notable feature is that the $\kappa$ becomes lighter 
than the $a_{0}$ meson, and the $\eta'$ meson lighter than 
the kaon, at about $240$ MeV.

The non-strange and the strange condensates, 
$\varphi_{\rm ns}$ and $\varphi_{\rm s}$,
are shown in Fig.\ \ref{fig3}(c). Note that 
the strange condensate decreases much more slowly 
with temperature than the non-strange condensate. 
Figure \ref{fig3}(d) shows the scalar and pseudoscalar 
mixing angles.  At large temperatures, these approach 
$\arcsin (1/\sqrt{3}) \simeq 35.264^{\rm o}$. From Eq.\ (\ref{ortho}), 
\begin{mathletters}
\bea
\sigma \equiv 
        \tilde{\sigma}_{0} &=& \sqrt{\frac{2}{3}} \, \sigma_{0} + 
        \frac{1}{\sqrt{3}} \, \sigma_{8}  \,\, , \\ 
f_{0} \equiv
        \tilde{\sigma}_{8} &=& -\frac{1}{\sqrt{3}} \, \sigma_{0} + 
        \sqrt{\frac{2}{3}} \, \sigma_{8} \,\, .
\eea
\end{mathletters}
On the other hand, 
\begin{mathletters}
\bea
\varphi_{\rm ns} &=& \sqrt{\frac{2}{3}} \, 
        \bar{\sigma}_{0} + 
        \frac{1}{\sqrt{3}} \, \bar{\sigma}_{8}  \,\, , \\ 
\varphi_{\rm s} 
        &=& \frac{1}{\sqrt{3}} \, \bar{\sigma}_{0} - 
        \sqrt{\frac{2}{3}} \, \bar{\sigma}_{8} \,\, .
\eea
\end{mathletters}
This shows that the $\sigma$ meson becomes an excitation 
of the non-strange condensate, i.e., a purely non-strange
$q \bar{q}$ state.  On the other hand, the $f_{0}$ meson 
is an excitation of the strange condensate and a pure 
$s \bar{s}$ state. 
Similarly, the $\eta'$ meson is purely non-strange and the $\eta$ meson
is purely strange. 
This is what was referred to as ideal flavor mixing earlier.

The transition occurs at temperatures which are not significantly
larger than the strange quark mass. Therefore, the explicit 
$SU(3)_r \times SU(3)_\ell$ symmetry breaking by the 
strange quark mass cannot be neglected and, 
at first, only the (approximate) $SU(2)_r \times SU(2)_\ell$ symmetry is
restored. This means that the pion becomes degenerate with
the $\sigma$ meson, and the $a_0$ becomes degenerate with the 
$\eta'$ meson. (The $\eta'$ is purely non-strange due to ideal
flavor mixing.) However, due to the fact that the
strange condensate decreases rather slowly with temperature, 
the explicit $U(1)_A$ symmetry breaking term $\sim {\cal G}_{abc}
\bar{\sigma}_c$ in Eqs.\ (\ref{massmatrices2}) is not small.
This causes the pion/$\sigma$ meson mass still to be different from 
the $a_0/\eta'$ meson mass.
Only when both condensates approach zero, the $U(1)_A$ symmetry is
effectively restored and the masses of all non-strange particles
become degenerate.

When the temperature becomes significantly larger than $m_s$,
the (approximate) $SU(3)_r \times SU(3)_\ell$ symmetry 
is restored. Then, all scalar octet states become degenerate, likewise
all pseudoscalar octet states become degenerate. 
If this happened when the explicit $U(1)_A$ breaking
term was still large, then the complete pseudoscalar octet would become
degenerate in mass with the scalar singlet, and the scalar octet 
degenerate in mass with the pseudoscalar singlet.
As it turns out, however, the explicit $U(1)_A$ symmetry breaking
becomes small around the same point where the (approximate) 
$SU(3)_r \times SU(3)_\ell$ symmetry is restored.

\vspace*{0.5cm}

\subsection{Explicit chiral symmetry breaking without
$U(1)_A$ anomaly}

\begin{figure}
\hspace*{1cm}
\mbox{\epsfig{file=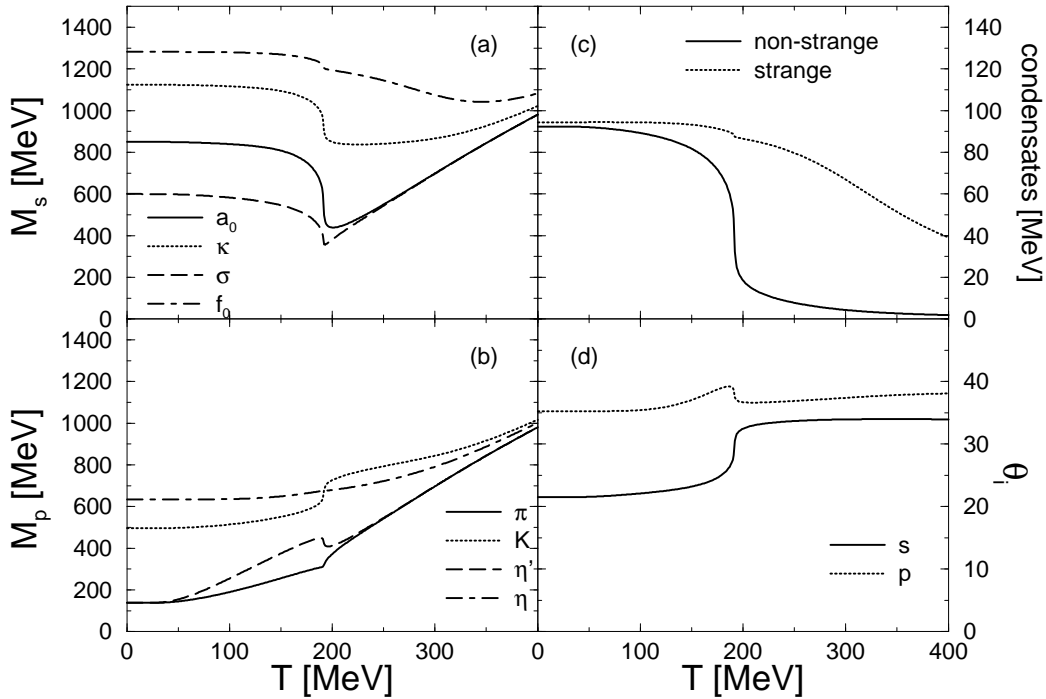,height=8cm,angle=270}}
\vspace{0cm}
\caption{As in Fig.\ \ref{fig3}, but for $c=0$.\label{fig4}}
\end{figure}

In Fig.\ \ref{fig4}, we show the masses for the scalars (a), 
the pseudoscalars (b), the condensates (c), and the mixing 
angles (d) for explicit chiral symmetry breaking in the
absence of the $U(1)_{A}$ anomaly.  
The masses behave according to the 
discussion of case 4(b) in Sec.\ \ref{III}.  As the temperature 
increases, the chiral symmetry restoration 
crossover transition is much more rapid than in 
the previous case, and occurs at a slightly 
smaller temperature, $\sim 200$ MeV.
A notable feature is the inverse
mass ordering of the $\eta$ meson and the kaon. 
At small temperatures and above the transition, the 
masses of the pion and the $\eta'$ meson are the same.  In the 
temperature range from about 50 to 210 MeV, however, they 
differ. We perceive this to be an artefact of the 
violation of Goldstone's theorem in the Hartree approximation, 
cf.\ Fig.\ \ref{fig2}(c).  

The melting of the condensates, Fig.\ \ref{fig4}(c), is 
similar to the previous case,  Fig.\ \ref{fig3}(c).  The mixing angles, 
Fig.\ \ref{fig4}(d), again approach ideal flavor mixing 
at large temperatures.  The difference here, however, is 
that the $\eta$ and $\eta'$ mesons are also ideally flavor-mixed 
at zero temperature, cf.\ Sec.\ \ref{IVb}.

Due to the absence of the $U(1)_A$ anomaly, once
the (approximate) $SU(2)_r \times SU(2)_\ell$ symmetry is
restored, the pion, the $\eta'$, the $a_0$, and the $\sigma$
mesons simultaneously become degenerate in mass. (The $\eta'$ meson
belongs to this class of non-strange particles due to ideal
flavor mixing.) Once the temperature becomes large compared
to the strange quark mass, the masses of the strange mesons
converge with those of the non-strange mesons.

\vspace*{0.5cm}

\section{Conclusions}\label{VII}

In this work, we computed properties of the $SU(3)_{r} \times 
SU(3)_{\ell}$ linear sigma model in the Hartree approximation 
at nonzero temperature.  We first classified 
possible patterns of symmetry breaking, with 
special attention to the cases where 
the $U(1)_{A}$ anomaly is either absent or present, 
and the cases of zero, degenerate or
nonzero, non-degenerate quark masses.
We then determined the coupling constants 
from the vacuum values for the masses and the decay 
constants in the various cases of interest.
We systematically derived the 
Hartree approximation within the CJT formalism.
Within this approximation, we  
computed the masses of scalar and pseudoscalar particles, 
the non-strange and strange condensates, and the scalar 
and pseudoscalar mixing angles as a function of temperature.   
We checked that our results are consistent with the  
mean-field approximation employed in \cite{Jurgen1} to 
compute these quantities.

For the $SU(N_f)_{r} \times SU(N_f)_{\ell}$ model,
in the case where the quark masses are zero, universality 
arguments predict the chiral symmetry restoring transition to 
be first order for $N_f = 3$ and $N_f=2$ in the absence of
the $U(1)_A$ anomaly, and second order for 
$N_f = 2$ in the presence of the $U(1)_A$ anomaly \cite{PisarskiWilczek}.
We find that the Hartree approximation correctly gives 
a first order transition in the case $N_f=3$. 
This is not necessarily an indication for the validity 
of this approximation, because
earlier work has shown that 
it incorrectly predicts a first order 
transition when $N_f=2$ and the $U(1)_A$ anomaly is present \cite{JDRenorm}.
The transition temperature is on the order of 170 MeV.

As expected, when the $U(1)_{A}$ anomaly is absent, 
the $\eta'$ meson becomes a Goldstone boson
for zero quark masses.  A surprising result is that then 
the $\sigma$ meson mass is directly proportional to the 
condensate.  

For nonzero quark masses, $m_{u}=m_{d} \neq m_{s}$,  
we find the transition to be a crossover transition, 
but for vanishing $U(1)_{A}$ anomaly the crossover 
region is much more narrow than in the presence of the
$U(1)_{A}$ anomaly.
In the chirally symmetric phase, the mixing angles 
approach the situation of ideal flavor mixing, i.e., 
the $\sigma$ and $\eta'$ mesons are pure non-strange 
$q \bar{q}$ states, while the $f_0$ and the $\eta$ mesons 
are pure $s \bar{s}$ states.
   
As an outlook, the present framework can be used 
as an alternative to lattice QCD studies \cite{Columbia}
to study the order of the chiral symmetry restoring 
transition as a function of the strange and non-strange 
quark masses \cite{jtlorder}.  Moreover, other meson 
properties such as the decay widths and the spectral 
functions can be self-consistently computed at 
nonzero temperature \cite{jj}. These properties can be experimentally
investigated in relativistic nuclear collisions, for instance
at Brookhaven National Laboratory's Relativistic Heavy-Ion Collider.

\begin{center} 
{\bf Acknowledgements} 
\end{center}

We thank T.\ Blum, D.E.\ Kharzeev, and R.D.\ Pisarski
for valuable discussions.
D.H.R.\ and J.S.-B.\ thank RIKEN, BNL and the U.S.\ Dept.\ of Energy for
providing the facilities essential for the completion of this work.
D.H.R.\ thanks Columbia University's Nuclear Theory group for continuing
access to their computing facilities.
J.T.L.\ thanks BNL's Nuclear Theory group for
support and hospitality during the completion of 
this work.  J.T.L.\ is supported by the Director, Office of Energy
Research, Division of Nuclear Physics of the Office of 
High Energy and Nuclear Physics of the U.S.\ Department of 
Energy under Contract No.\ DE-AC02-98CH10886.

\appendix

\section*{Derivation of Eqs.\ (\ref{SIGMA_08})}

The infinitesimal form of the  $SU(3)_{r} \times SU(3)_{\ell}
\times U(1)_{A}$ symmetry transformation (\ref{trans}) is 
\be 
T^{a} \, \phi^{a} \longrightarrow T^{a} \, \phi^{a}  - 
        i \, \omega_{V}^{a} \,  \left[ T^{a}, T^{b} 
        \right] \, \phi^{b} + i\, 
        \omega_{A}^{a} \, \left\{ T^{a}, T^{b} 
        \right\} \, \phi^{b} \,\, .
\ee
For axial-vector transformations, $\omega_{V}^{a} \equiv 0$, and 
the associated (axial-vector) Noether current is 
\bea
{\cal J}^{\mu}_{a} &\equiv& \frac{\delta \, {\cal L}}{\delta \left(
        \partial_{\mu} \phi_{b} \right) } 
        \, i \, d_{abc} \, \phi_{c}  + {\rm h.c.} \nonumber \\
        &=& \frac{i}{2} \, \left( \partial^{\mu} \sigma_{b} - 
        i \, \partial^{\mu} \pi_{b} \right) \, d_{abc} \, 
        \left( \sigma_{c} + i \, \pi_{c} \right) + {\rm h.c.} \nonumber \\
        &=& d_{abc} \left( \sigma_{b} \, \partial^{\mu} \pi_{c}  
        - \pi_{b} \, \partial^{\mu} \sigma_{c}  
        \right) \,\, .  
\eea
Inserting this into the PCAC relation, 
\be
\langle 0| {\cal J}^{\mu}_{a} | \pi_{a} \rangle \equiv 
        i \, p^{\mu} \, f_a \,\, , 
\ee 
where $f_{a}$ is the decay constant corresponding to the 
field $\pi_{a}$, and shifting the scalar fields by their 
vacuum expectation values, $\sigma_{a} \rightarrow 
\sigma_{a} + \bar{\sigma}_{a}$, one obtains  
\be
f_{a} = d_{aab} \, \bar{\sigma}_{b} \,\, ,
\ee
where one sums over the index $b$ but not over $a$.

In the case that $\bar{\sigma}_{0}$, 
$\bar{\sigma}_{8} \neq 0$, one obtains for the pion and kaon decay 
constants
\begin{mathletters}
\bea
f_{\pi} &\equiv& f_{1} = d_{11a} \, \bar{\sigma}_{a} = 
        \sqrt{\frac{2}{3}} \, \bar{\sigma}_{0} + 
        \frac{1}{\sqrt{3}} \,  \bar{\sigma}_{8} \,\, , \\
f_{K} &\equiv& f_{4} = d_{44a} \, \bar{\sigma}_{a} = 
        \sqrt{\frac{2}{3}} \, \bar{\sigma}_{0} - 
        \frac{1}{\sqrt{12}} \,  \bar{\sigma}_{8} \,\, .
\eea
\end{mathletters}
In the case that $\bar{\sigma}_{8} = 0$, this simplifies to 
$f_{\pi} = f_K = \sqrt{2/3} \, \bar{\sigma}_{0}$.

\end{document}